 \documentclass[iop]{emulateapj}

\usepackage{epsfig}          
\usepackage{graphicx}        
\usepackage{amssymb}        
\usepackage{url}             
\usepackage{float}
\usepackage{lineno}

\newcommand{\etal}{{\it et al.}}

\newcommand{\be}{\begin{equation}}
\newcommand{\ee}{\end{equation}}
\newcommand{\bea}{\begin{eqnarray}}
\newcommand{\eea}{\end{eqnarray}}
\newcommand{\ba}{\begin{array}}
\newcommand{\ea}{\end{array}}

\newcommand{\bra}{\langle}
\newcommand{\ket}{\rangle}
\newcommand{\bit}{\begin{itemize}}
\newcommand{\eit}{\end{itemize}}
\newcommand{\ben}{\begin{enumerate}}
\newcommand{\een}{\end{enumerate}}

\begin{document}

\title{Meridional circulation dynamics from 3D MHD global simulations of solar convection}

\author{D\'{a}rio Passos \altaffilmark{1,2,3},
        Paul Charbonneau \altaffilmark{2},
        Mark Miesch \altaffilmark{4}}

\altaffiltext{1}{CENTRA, Instituto Superior Tecnico, Universidade de Lisboa,
                 Av. Rovisco Pais 1, 1049-001 Lisbon, Portugal ;
                 Email: dariopassos@ist.utl.pt}

\altaffiltext{2}{D\'epartment de Physique, Universit\'e de Montr\'eal,
                 C.P. 6128, Centre-ville, Montr\'eal, Qc, Canada  H3C-3J7}

\altaffiltext{3}{Departamento de F\'isica da Universidade do Algarve,
                 Campus de Gambelas, 8005-139 Faro, Portugal}

\altaffiltext{4}{High Altitude Observatory, NCAR, Boulder CO 80301-2252, USA}

\begin{abstract}

The form of the solar meridional circulation is a very important ingredient
for mean field flux transport dynamo models. Yet a shroud of mystery still surrounds
this large-scale flow, given that its measurement using current helioseismic
techniques is challenging.
In this work we use results from 3D global simulations of solar convection
to infer the dynamical behavior of the established meridional circulation. We make a direct
comparison between the meridional circulation that arises in these simulations and the
latest observations. Based on our results we argue that there should be an equatorward
flow at the base of the convection zone at mid latitudes, below the current maximum
depth helioseismic measures can probe (0.75 R$_\odot$). We also provide physical
arguments to justify this behaviour. The simulations indicate that the meridional
circulation undergoes substantial changes in morphology as the magnetic cycle
unfolds.
We close by discussing the importance of these dynamical changes for current
methods of observation that involve long averaging periods of helioseismic data.
Also noteworthy is the fact that these topological changes indicate a rich
interaction between magnetic fields and plasma flows, which challenges the
ubiquitous kinematic approach used in the vast majority of mean field
dynamo simulations.

\end{abstract}



\section{Introduction}
The pressing need for a better understanding of space weather phenomenology and its
implications has provided impetus for ongoing improvements in solar dynamo models.
Nonetheless a fully detailed and consensual picture adequately explaining
the origin and
evolution of the large-scale solar magnetic field has not yet materialized.
Amongst the several types of extant mean-field models, \citep[see review by][]{Charbonneau2010},
a specific class of dynamo models known as
mean-field Babcock-Leighton Flux Transport Dynamos (FTD) has been particularly
successful in explaining some of the main observational features of the
solar cycle, e.g. \citep{Wang1991, Dikpati1999, Nandy2002, Karak2014}.
These models rely heavily on the advective role of the meridional
circulation (MC), a relatively weak, large-scale plasma flow that its
thought to act as a conveyor belt in the convection zone (CZ),
transporting magnetic flux in radius and latitude.
In these models (running mainly in the kinematic regime),
the MC is an important determinant of the period and amplitude
of the cycle \citep[][and references therein]{Charbonneau2010}.

From an observational point of view, the meridional circulation is known with a good
degree of confidence at low to mid latitudes in the photospheric layers of the Sun,
through Doppler measurements and tracking of surface features
\citep{Ulrich2010,Hatha2012}.
This surface component of the MC appears to play an important role in the inversion
of the solar dipole. Observations and surface advection models suggest that it
effectively transports the residual field produced by the decay of active regions
(the Babcock-Leighton mechanism) towards the poles, contributing to the polarity
reversal \citep{Babcock1961, Cameron2011, MunozJaramillo2013}.
However, in the solar interior, due to the weak nature of this flow,
its measurement is experimentally challenging and a complete profile mapping over the whole
CZ has not yet been achieved by helioseismology.
Since until recently the observational constraints on the MC profile were
restricted to the surface layers of the Sun, FTD models employed
extrapolations to the solar interior based on mass conservation and
simplified MC profiles with one or two cells per hemisphere.
Of particular importance is the so-called \textit{return flow}, an equatorward
directed component of the MC at the base of the CZ that,
in the context of FTD, is responsible for the equatorward migration of the dynamo
wave over the course of the cycle \citep{Wang1991,Chou1995}. Even if more complex
MC topologies are assumed, this equatorward flow at the base of the CZ retains
a preponderant role \citep{HazraG2014}.

Recently, helioseismic measurements by \cite{Zhao2013} mapped the MC
down to a depth of 0.75 R$_\odot$ for latitudes below 60$^\circ$.
Based on their measurement they then
extrapolate the behavior of the MC all the way down to the base of the CZ
and suggest that this large-scale flow is globally structured as an upper cell that
rotates counter-clockwise (in the northern hemisphere)
and a bottom cell that rotates clockwise. This suggested configuration
implies that the flow direction at the base of CZ is poleward directed.
See figures 1 and 4a of \cite{Zhao2013}.
However, when this proposed MC cell topology is used in FTD models, solar-like
dynamo solutions are no longer produced (see section 3 of \cite{HazraG2014}).
Though these new observational developments are potentially transformative, we must
emphasize that the actual measurements place little constraint on the amplitude and direction
of the flow near the base of the CZ, where it matters most for regulating the cycle in FTD.

Another interesting result in need of a theoretical
interpretation is the observed variation of the surface MC, the amplitude
of which varies in anti-phase with the solar cycle (Komm et al. 2009, Hathaway 2010).
This and the development of a high-latitude counter-rotating secondary cell in the
descending phase of the cycle \citep{Jiang2009,Ulrich2010}, hints at possible
changes in the overall topology of the MC over the course of the solar cycle. Understanding
how this process occurs requires modelling techniques that
are dynamically consistent.

Unlike kinematic dynamo models in which the flows are prescribed and
their consequences studied, global convective dynamo simulations self-consistently
establish the meridional and zonal flows as a consequence of the
internal fluid dynamics, \citep[e.g.][]{Clune1999, Brun2004, Ghizaru2010, Kapyla2012}.
In this work we use one such 3D MHD simulation of the solar CZ to
explore the topological structure of the MC and its implications
within the context of magnetic cycles, helioseismic measurements and
mean-field kinematic dynamo simulations.

The data used in this study is obtained from an Implicit Large-Eddy Simulation (ILES)
of global solar convection produced by the EULAG-MHD code \citep{SmolarCharb2013}, more
specifically the \textit{millennium simulation} recently presented in
\cite{PassosCharb2014} (henceforth PC14) and based on a setup described
in \cite{Ghizaru2010} and \cite{Racine2011}. This simulation is performed
on a 128$\times$64$\times$47 mesh in ($\phi,\theta,r$) which allows long integration times.
The numerical dissipation introduced by the ILES scheme amounts to an implicit adaptive subgrid
model which maintains numerical stability. In contrast to DNS, information on small-scale
dissipation is not readily retrievable, but the resolved scales are dynamically consistent.
These simulations reach a turbulent regime characterized by considerable
dynamic range when zonal averages are used to define large scales and
fluctuations around these averages are used to define small scales, \citep{Racine2011}.

\section{Meridional circulation structure}

\begin{figure*}[htb!]
\centering
\includegraphics[width=170mm]{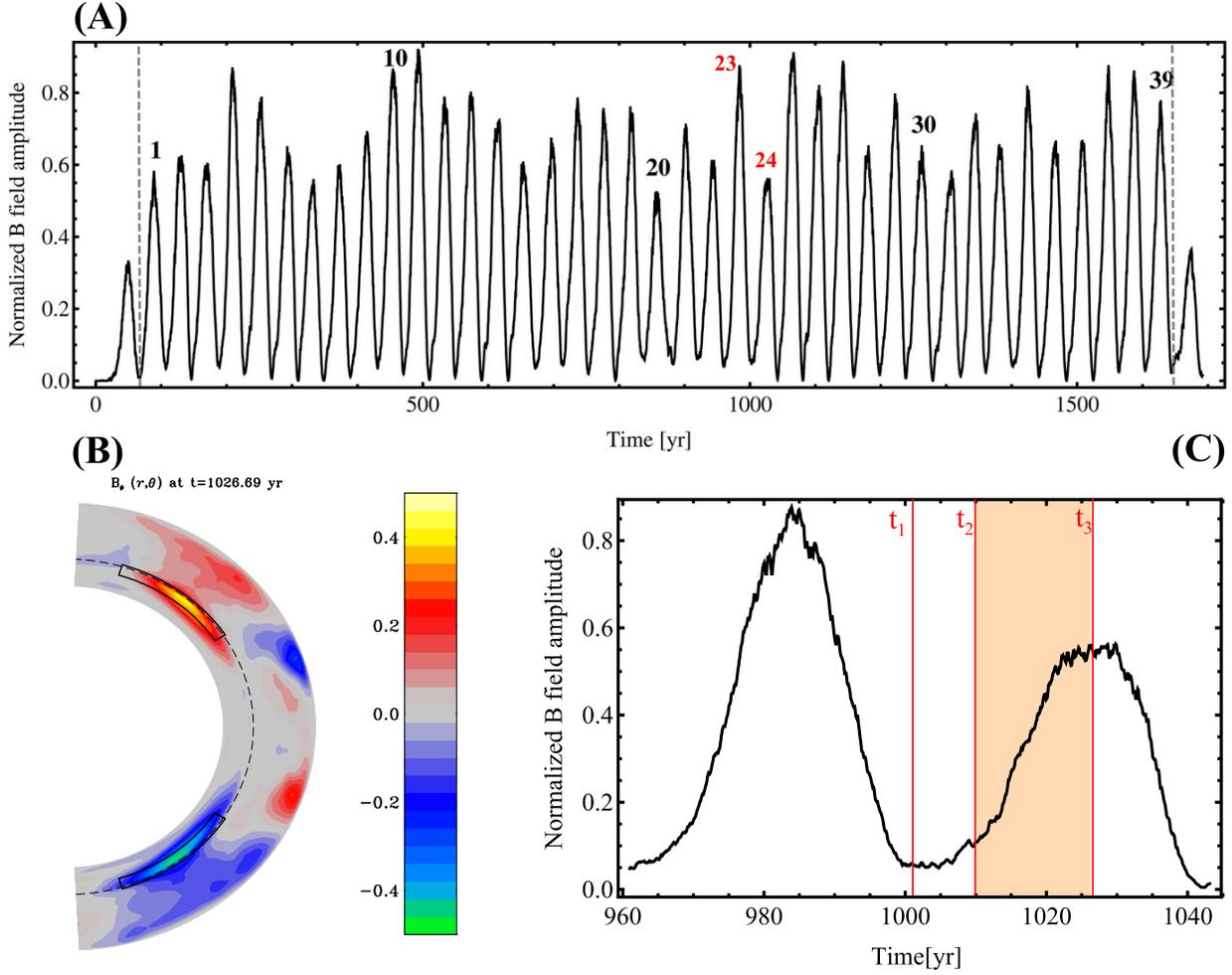}
\caption{(A) full disk toroidal field proxy, $B$. (B) Meridional representation
of the toroidal field at the maximum of simulated cycle 24 taken at t$_3$. The boxed
areas mark the integration domains used to build the toroidal proxy. (C) its
a zoom in on cycle 23 and 24 of this simulation. The orange rectangle indicates
the averaging period over which $u_\theta$ is computed. Compare (C) to Fig.~4
in \cite{Zhao2013}.}
\label{fig:proxyB}
\end{figure*}

Before we investigate the dynamics of the MC in our simulations, we
first assess the degree of resemblance between the simulated
meridional flow and the solar MC.
In \cite{Zhao2013} the authors use helioseismic measurements carried out by SDO
(HMI Doppler-shift data) averaged over a two year period, spanning from
May 1, 2010 to April 30, 2012 which corresponds to the rising phase of
sunspot cycle 24. For our initial assessment we will use a time window corresponding
to the same phase of the magnetic cycle in our simulation.
To represent solar magnetic activity we built a proxy for the large-scale
magnetic field by integrating the zonally averaged toroidal field $\langle B_\phi \rangle$
over an extended region centered at the depth and latitudes where the field
accumulates (see Fig. \ref{fig:proxyB}b), following the procedure described in
detail in PC14.
This toroidal proxy is our analog of the sunspot number.
The large-scale magnetic cycle developing in this simulation shows
regular polarity reversals with a period on the order of 40 years, with the
large-scale magnetic field peaking at the base of the convection zone at
mid/high latitudes (see figures 1 and 2 of PC14).

In order to produce a latitudinal velocity diagram ($u_\theta$) comparable to the
one presented in figure 4 of \cite{Zhao2013}, we choose a cycle from the simulation
with similar characteristics as sunspot cycle 24, i.e. broadly speaking a low
amplitude cycle following one of higher amplitude.
Coincidentally, the 24$^{\rm th}$ cycle in our millennium simulation is a
good candidate to perform this comparison.
Figure \ref{fig:proxyB}C shows the period over which we averaged our signal, i.e. over the
rising phase of the cycle which is analogous to the 2 year interval used in \cite{Zhao2013}
(nearly 20\% of the cycle duration in both cases).

\begin{figure*}[htb!]
\centering
\includegraphics[width=170mm]{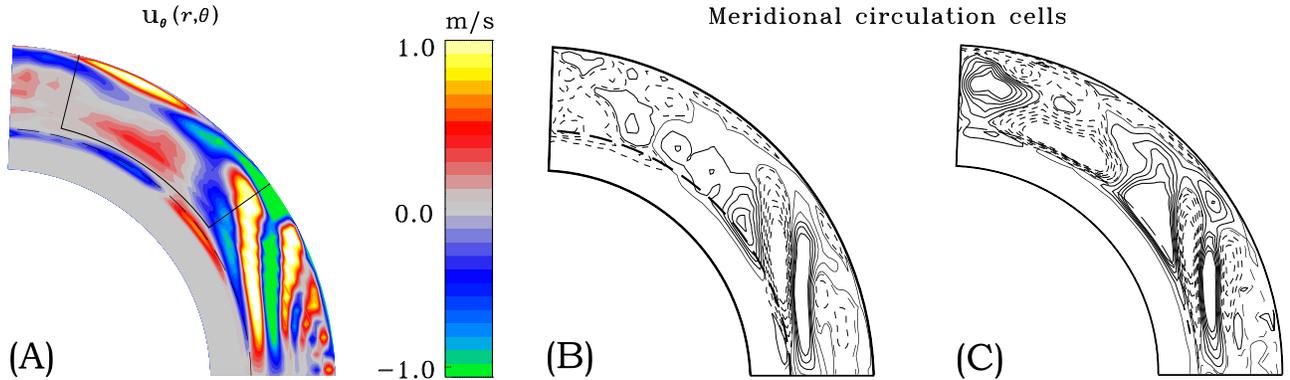}
\caption{(A) large-scale $\bra u_\theta \ket$ component of the meridional
flow averaged over the rising phase of simulated cycle 24 in the northern
hemisphere. The dashed line represents the tachocline. Positive (negative)
velocity values indicate poleward (equatorward) flows.
The corresponding stream function profile that
indicates the cell structure is plotted for (B) cycle minimum (at $t_1$
on Fig.~\ref{fig:proxyB}C) and (C) cycle
maximum (at $t_3$). Dashed (solid) contours in (B) and (C)
represent counter-clockwise (clockwise) rotation.}
\label{fig:mc1}
\end{figure*}

The large-scale axisymmetric latitudinal velocity component, $\langle u_\theta \rangle$,
is extracted from the simulation also by zonal averaging (see Fig. \ref{fig:mc1}A).
Here we follow the convention usually used in EULAG-MHD, i.e. $u_\theta$ is defined
positive towards the poles.
Using $\langle u_r \rangle$ and $\langle u_\theta \rangle$, in a similar
manner to that presented in \cite{Guerrero2013}, we also compute the stream function
of the meridional flow which is useful for understanding the cell morphology
(see panels (B) and (C) of figure \ref{fig:mc1}).

A detailed comparison of the helioseismically-inferred internal meridional
flow with our EULAG-MHD simulation is complicated by the fact that the
large-scale magnetic cycle developing in the latter peaks at much higher
latitude ($\sim 55^\circ$) than the sunspot butterfly diagram ($\sim 15^\circ$),
and shows only hints of equatorial migration.
The large-scale field migration toward lower latitudes in the simulation
appears to be inhibited by the presence of rotationally-aligned ``banana cells''
outside of a cylinder tangent to the equatorial base of the convection zone.
The existence of these cells reflects the strong influence of rotation on
turbulent convection, and is a robust feature observed in all numerical simulations
of rotating convection in the low Rossby number regime, with or without magnetic
fields (see, e.g., Miesch \& Toomre 2009, \S 2.2.3; also Guerrero et al.~2013).
However, the dynamical response of the simulation to the magnetic cycle
poleward of the simulated ``activity belts'' shows
some encouraging similarities to the sun.
For example, Beaudoin et al. (2012) have shown that the pattern
of cyclic rotational torsional oscillations developing poleward of the
mid-latitude regions of peak large-scale magnetic fields show the same
amplitude, phasing, and even double-branch structure as inferred
helioseismically (see their Fig.~5 and accompanying discussion).
This offers hope that the meridional circulation dynamics
may also be similar, considering that meridional and zonal force balances
are tightly coupled (more on this in \S 3 herein).

\begin{figure*}[htb]
        \centering
        \includegraphics[width=16 cm]{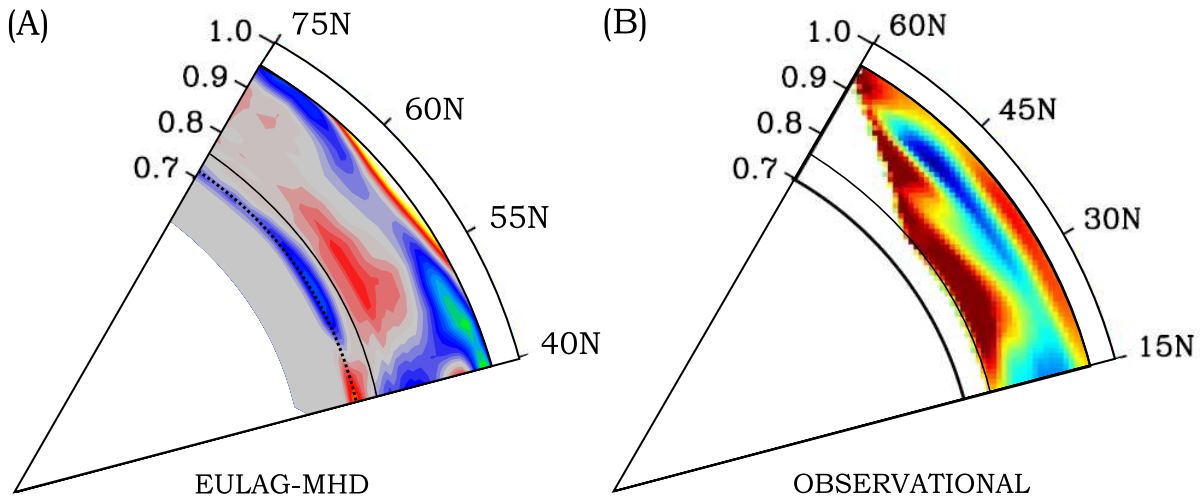}
        \caption{(A) $\bra u_\theta \ket$  profile extracted from the
        \textit{millennium simulation} between 40 and 75 degrees north. The
        dotted line represents the tachocline depth.
        (B) $u_\theta$  profile measured by \cite{Zhao2013} between 15 and
        60 degrees north. The solid black line at 0.75 R$_\odot$ represents
        the maximum depth probed in the measurement.
        The base color scheme is the same, namely warm color correspond to
        poleward flows and cold, a comparison between both images show a
        good resemblance in terms of the topology of this flow.
        An equatorward return flow (in blue) is clearly visible
        at tachocline depths in (A), a depth not yet reached by helioseismic
        measurements.}
        \label{fig:eulag-zhao}
\end{figure*}

Figure \ref{fig:eulag-zhao} replots the mid- to high-latitude
portion of the $\langle u_\theta \rangle$ profile in Fig.~\ref{fig:mc1}A,
artificially remapped to the 15---60$^\circ$ latitude range of the
\cite{Zhao2013} data, the latter plotted in (B) for comparison.
A key feature of the MC extracted from the simulation
is an equatorward flow present below the largest depth
probed by
\cite{Zhao2013}, indicated by the solid circular arc in the left
panel of figure \ref{fig:eulag-zhao}. This equatorward
flow is part of a counter-clockwise meridional cell appearing at those latitudes
and depths.
Panels (B) and (C) of figure 2 show the different morphological configurations
of the MC profile at cycle minimum and cycle maximum respectively. The cell where the
return flow is located is much more pronounced at cycle maximum. Moreover,
as originally reported in \cite{Passos2012}, the amplitude of this equatorward
flow varies in phase with the build up of the large-scale toroidal. This is an
indication that the MC at this depth is highly dynamical and magnetically influenced.
 Another indication of this is the fact that the MC in
this simulation is stronger than that arising in an analogous simulation using
a purely hydrodynamic set up (no magnetic fields).

\section{Meridional circulation dynamics}

The existence of the equatorward flow at mid-latitudes near the base of the convection
zone is a robust feature of global convection simulations, and particularly those that
include a stable radiative zone below the CZ (Miesch 2005).  This generally occurs
even if the MC in the bulk of the CZ is multi-celled. It can be attributed to two
distinct but complementary physical mechanisms, namely gyroscopic pumping and
turbulent alignment of convective plumes.  Both are essentially hydrodynamic
mechanisms but magnetic feedbacks can alter the dynamical balances, inducing
time variations.

Gyroscopic pumping refers to the tendency for a zonal force (axial torque)
to induce a
meridional flow through the inertia of the differential rotation
\citep{McIntyre2000, Miesch2011}.
In convection simulations the zonal force arises principally from the
convective Reynolds stress, which tends to transport angular momentum
radially inward at mid-latitudes ($\left<u_r^\prime u_\phi^\prime\right> < 0$)
due to the Coriolis deflection of downflow plumes and broader upflows
(Featherstone \& Miesch 2014).  The convergence of the inward angular momentum
flux at the base of the CZ acts to accelerate the zonal flow and induces an
equatorward flow that acts to offset this acceleration.  Note that this
convergence of the convective angular momentum flux near the base of the
CZ does not preclude one or more other regions of convergence closer to
the surface.  In particular, a similar convergence of inward angular
momentum flux at the base of the near-surface shear layer due to plumes driven
in the photospheric boundary layer may account for the shallow return
flows inferred from recent photospheric and helioseismic measurements
(Hathaway 2012; Zhao et al.\ 2013).  In other words, the presence of an
equatorward return flow at $\sim 0.95\,$R$_\odot$ does not necessarily
rule out an equatorward flow at the base of the CZ as well.

Turbulent alignment refers to the tendency for helical downflow plumes to
be diverted away from the vertical direction and toward the rotation axis
($\left<u_r^\prime u_\theta^\prime\right> > 0$ in the northern hemisphere,
opposite in the southern) due to the Coriolis deflection of flow components
that are perpendicular to the rotation axis (Brummell et al.\ 1996, 1998).
As these plumes enter the stably stratified fluid layers
their vertical momentum is halted by the buoyancy force but they retain their
equatorward momentum.  The cumulative effect of many such plumes can produce
a net equatorward circulation in the overshoot region and lower CZ (Miesch 2005).

\begin{figure*}[htb!]
\centering
\includegraphics[width=170mm]{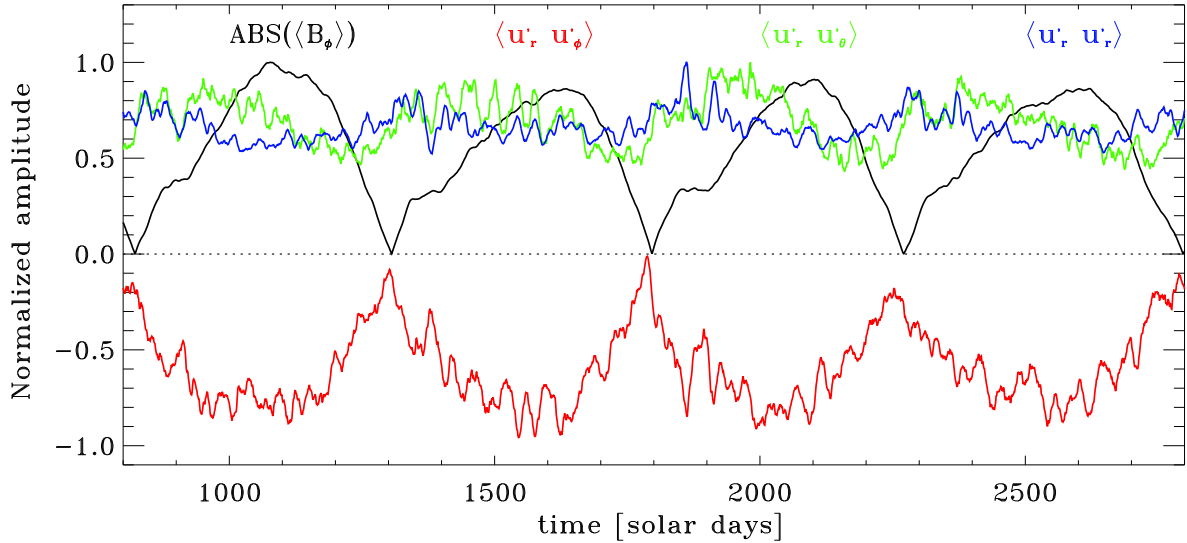}
\caption{Time series of
         of the axisymmetric
         toroidal field, $\langle B_\phi \rangle$ (in black), and of the Reynolds
         stress components
         $\langle u'_r\, u'_\phi \rangle$, $\langle u'_r\, u'_\theta \rangle$
         and $\langle u'_r\, u'_r \rangle$, in red, green and blue, respectively.
         All time series are computed by integration over the boxed region
         of Fig. \ref{fig:proxyB}b in the northern hemisphere only, and are
         all normalized to the highest value found in the two full magnetic cycles
         sampled.}
\label{fig:perturb_corr}
\end{figure*}

These two mechanisms for maintaining an equatorward flow near the base of the
CZ have been found previously in global convection simulations run with the
ASH code (Miesch \textit{et al.} 2000; Featherstone \& Miesch 2014).
Figure \ref{fig:perturb_corr} demonstrates that similar mechanisms
may be operating in our EULAG-MHD simulations. Shown are correlations among the
fluctuating velocity components defined as $u'_i= u_i - \langle u_i \rangle$,
where $u_i$ is the velocity component before zonal averaging and
$\langle u_i \rangle$ after zonal averaging.
These correlations were computed for the northern hemisphere and for 2
full magnetic cycles spanning 5 polarity reversals (enough for illustrative purposes)
over the integration region shown in Fig. \ref{fig:proxyB}B. This region
encompasses the
interface between convectively stable and unstable fluid layers.
The $\langle u_r^\prime u_\phi^\prime \rangle$ correlation is robustly negative,
consistent with the maintenance of an equatorward circulation by the convergence
of an inward angular momentum flux; the mechanism of gyroscopic pumping as
discussed above.  Meanwhile, the $\langle u_r^\prime u_\theta^\prime \rangle$
correlation is positive,
indicating turbulent alignment.  Both show variation with the phase of the
magnetic cycle, with the inward angular momentum flux particularly prominent
during cycle maximum.

Though these Reynolds stresses tend to induce an equatorward flow near the base
of the CZ, they are opposed at mid-latitudes by the Lorentz force.  Here the
cyclic generation of strong toroidal fields via the $\Omega$-effect extracts
energy from the rotational shear, decelerating lower latitudes and accelerating
higher latitudes. In the region of deceleration on the equatorward edge of the
toroidal bands, gyroscopic pumping induces a poleward meridional flow that
varies in strength as the bands are alternately generated and destroyed
(Fig.\ \ref{fig:mc1}B, C).  This establishes a marked clockwise circulation cell
at mid latitudes that extends into the mid-CZ.  At higher and lower latitudes
the convective angular momentum transport maintains a counter-clockwise cell
with equatorward flow at the base of the CZ. This high-latitude cell
largely disappears during cycle minima while the mid-latitude cell extends
poleward (Fig.\ \ref{fig:mc1}B).

The $\langle u'_r u'_r \rangle$ correlation in Fig. \ref{fig:perturb_corr}, which is
a measure of the convective overshoot,
is higher when the toroidal field is lower and vice
versa. This means that at this depth the penetration power of the
downflows is higher when the toroidal field is weaker.
At cycle maximum, when the amplitude of the toroidal
field is maximum, the overshoot layers just below the tachocline develops
a higher ``rigidity'', presumably due to the magnetic tension force,
and the downflows do not penetrate as deep. As a consequence, the overshooting downflows
deposit their latitudinal momentum within a thinner layer immediately beneath the bottom
of the CZ, thus driving a larger local acceleration of the equatorward flow than at cycle
minimum. The cyclic modulation observed in the fluctuating
flow correlations is an
important reminder that other forces are acting. We are currently
expanding our analysis of these results in order to understand the physical mechanisms
through which the large-scale magnetic field is modulating the turbulence.

\section{Conclusions}

Extant helioseismic measurements are now mapping the solar meridional circulation to
a depth of 0.75 R$_\odot$.
Based on 3D MHD global simulations of solar convection, we presented arguments
that favor the existence of an equatorward flow below 0.75 R$_\odot$, near the base
of the CZ at mid to high latitudes. This information is important for the
continuous development of FTD models, and offer a challenge to future helioseismic
measurements which will eventually refute or confirm this result.

The horizontal (latitudinal) component of the MC extracted from this simulation,
$u_\theta$, is compared to the latest MC measurements of
\cite{Zhao2013} focusing on the latitude range poleward of the activity belts
in both cases.
In the simulation, solar-like features (e.g. regular
large-scale magnetic cycle and torsional
oscillations) appear at higher latitudes than in the Sun, and we argue that
this is possibly due to the influence of the dynamics associated with
rotationally-aligned, persistent ``banana cells'' that are typically present
in this type of simulations.
In an appropriately remapped latitude range and over the depths simulated
($0.602 \leq r/R \leq 0.96$) our $u_\theta$ profile shows the same broad
characteristics as \cite{Zhao2013} mapping presented in Fig. \ref{fig:eulag-zhao}.
We also observe evidence that the deep equatorward flow strength and topology
is tightly connected
with the build up of the large-scale toroidal magnetic field. As a possible explanation,
we suggest that when this field component accumulates beneath the
tachocline the associated magnetic tension force decreases the penetration depth of
convective downflows. In turn, this process contributes to the enhancement of a
mid latitude equatorward flow at the base of the CZ by depositing angular momentum
at higher depths.

This is a very dynamic process that changes greatly over the course of the
solar cycle and is indicative of an important two-way dynamical
coupling between field and flow.
This result also suggests that special attention should be taken when interpreting the results
provided by helioseismology, which require long averaging periods to extract
the weak meridional flow from the background turbulence noise. We are led
to believe that long periods of data integration will dilute any dynamical
effects impacting the deeper meridional flow cells.
It should also be emphasized that this tight coupling between
field and flow should have an impact in kinematic simulations of
flux transport dynamo models.
The ubiquitous kinematic approach is clearly missing some
important aspects of the dynamical interactions between field and flow.
We are currently expanding the present work to better understand the nature
of the physical mechanism(s) dynamically modulating
the MC global topology in the presence of a large-scale magnetic cycle.

\begin{small}
D.Passos is thankful to Gustavo Guerrero for helping with the stream function
representation, to Sandra Braz for support, and acknowledges the financial support
from the Funda\c{c}\~{a}o para a Ci\^{e}ncia e Tecnologia grant
SFRH/BPD/68409/2010, POPH/FSE and GRPS-UdeM
and the University of the Algarve for providing office space.
We thank Kyle Auguston for helpful comments on the manuscript.
P.~Charbonneau is supported primarily by a Discovery Grant from
the Natural Sciences and Engineering Research Council of Canada. All
EULAG-MHD simulations reported upon in this paper were performed on the
computing infrastructures of Calcul Qu\'ebec, a member of the Compute Canada
consortium. The National Center for Atmospheric Research is sponsored by the
National Science Foundation.
\end{small}


\end{document}